\documentclass[a4paper,11pt]{article}
\usepackage[T1]{fontenc}
\usepackage[utf8]{inputenc}
\usepackage{lmodern}

\usepackage{amsmath}

\usepackage{graphicx}

\title{\textbf{Quintessence with Hybrid Potential}}
\author{\textbf{Soghra Tayfeh Bagheri}\\\small Institute of Theoretical Astrophysics, University of Oslo}

\begin{document}

\maketitle

\begin{abstract}
\ I present the numerical solution of equations of the evolution of a universe containing background fluids (radiation, dark matter and baryonic matter), plus a scalar matter field with a hybrid potential that is a combination of exponential potential and power-law potential. The plot of the evolution of density parameters is compatible with our universe; and today’s values of density parameters of dark energy, dark matter, baryonic matter, and Hubble parameter, and the age and size of our universe, found from this model, are very close to (and some times the same as) measured values.
 
\end{abstract}

\section{\small INTRODUCTION}
\ The existence of dark energy in the universe is hypothesized to explain the accelerated expansion of today's universe that is proved by observational data gathered from ground-based and space telescopes, from type Ia supernovas ; and the inflationary expansion of the very early universe is assumed to solve issues of the Big Bang theory, for instance, flatness problem, and isotropy and homogeneity of the universe in large scale.\\
 Cosmological constant (or vaccum energy density), is the first and simplest model of dark energy which has the same value everywhere in space for all the time. Cosmological constant is completely defined by one number, its magnitude[1]. But the value of energy density of vacuum, based on the result of different theories, is $ 10^{50}$   - $ 10^{120} $ times larger than the magnitude allowed by cosmology[2].\\
 Unlike the cosmological constant, quintessential models of inflation and dark energy, are dynamical, evolving component of universe, with possibility to be spatially inhomogeneous[3].\\
In the next section I will discuss about a quintessential scalar matter field with hybrid potential, which along with background fluids (radiation, dark matter and baryonic matter), gives a compatible model of our evolutionary universe.
\section{\small SCALAR MATTER FIELD WITH HYBRID POTENTIAL}
\ Energy density and pressure of a homogeneous scalar matter field $ \Psi  $ as defined in [4], are:
\begin{equation}
\rho _{\Psi }=\frac{1}{2}(1-\gamma ^{2}\kappa ^{2}\Psi ^{2}) \dot{\Psi }  ^{2}+V(\Psi ) 
\end{equation}
\begin{equation}
p_{\Psi }=\frac{1}{2}(1-\gamma ^{2}\kappa ^{2}\Psi ^{2}) \dot{\Psi }  ^{2}-V(\Psi ) 
\end{equation}
where $ \kappa ^{2}=8\pi G $, and $ \gamma  $ is a constant coefficient.\\
The first Friedmann equation for a spatially flat FRW model of universe filled with scalar field, $ \Psi  $ and background fluid, $ \rho _{bg} $, is:
\begin{equation}
 H^{2} =\frac{\kappa ^{2}}{3}\rho _{total}=\frac{\kappa ^{2}}{3}\left[\frac{1}{2}(1-\gamma ^{2}\kappa ^{2}\Psi ^{2}) \dot{\Psi }  ^{2}+\rho _{bg}  +V(\Psi )\right] 
\end{equation}
By defining new field $ u_{\mp } $ as [4]:
\begin{equation}
du_{-}=\sqrt{1-\gamma ^{2}\kappa ^{2}\Psi ^{2}}\ d\Psi \ ,\ \ \ for \ the\ region\ \Psi <\frac{1}{\gamma \kappa }
\end{equation}
\begin{equation}
du_{+}=\sqrt{\gamma ^{2}\kappa ^{2}\Psi ^{2}-1}\ d\Psi \ ,\ \ \ for \ the\ region\ \Psi >\frac{1}{\gamma \kappa }
\end{equation}
the first Friedmann equation will be simplified as:
\begin{equation}
 H^{2} =\frac{\kappa ^{2}}{3}\left[\pm \frac{1}{2}\dot{u}^{2}_{\mp }+\rho _{bg}  +V(\Psi )\right] 
\end{equation}
The hybrid potential, $ V(\Psi ) $ can be defined as:
\begin{equation}
V(\Psi )=V_{e}(\Psi )+V_{p}(\Psi )=\delta e^{-(\kappa \Psi )^{\beta }}+\lambda \kappa ^{\alpha -4}\Psi ^{\alpha }
\end{equation}
where $ \delta  $, $ \beta  $, $ \lambda  $ and $ \alpha  $, are constant parameters. 

\section{\small DIMENSIONLESS VARIABLES AND THE EVOLUTION OF THE UNIVERSE}
According to the Eq. (6), density parameters (or dimensionless variables) of a universe filled with scalar field $ \Psi  $ with hybrid potential and background fluids (radiation, dark and baryonic matter), are:\\
\begin{equation}
 x_{\mp }^{2}={\frac{\kappa ^{2}\dot{u}^{2}_{\mp } }{6H^{2}}}\ \ \ \ ,\ \ \ \ \ y_{e}^{2}={\frac{\kappa ^{2}V_{e}(\Psi  )}{3H^{2}}}\ \ \ \ ,\ \ \ \ \ y_{p}^{2}={\frac{\kappa ^{2}V_{p}(\Psi  )}{3H^{2}}} 
  \end{equation}
 \begin{equation}  
  r^{2}={\frac{\kappa ^{2}\rho _{r}}{3H^{2}}}\ \ \ \ \ \ \ ,\ \ \ \ \ \ m_{d}^{2}={\frac{\kappa ^{2}\rho _{d}}{3H^{2}}}\ \ \ \ ,\ \ \ \ m_{b}^{2}={\frac{\kappa ^{2}\rho _{b}}{3H^{2}}} 
 \end{equation} \\
 where $ \rho _{r} $, $ \rho _{d} $, and $ \rho _{b} $, are the densities of radiation, dark matter and baryonic matter, respectively. Density parameters are constrained by the first Friedmann equation, as:
 \begin{equation}
 \pm x^{2}_{\mp} +y^{2}_{e}+y^{2}_{p}+r^{2}+m_{d}^{2}+m_{b}^{2}=1
 \end{equation}
The derivatives of the dimensionless variables with respect to the logarithm of the scale factor, a, defined as: 
 \[ '=\frac{d}{d(Ln(a))}=\frac{d}{dN} \]
will give the evolution of our universe. But first we need to find the equations of motion of scalar field, $ u_{\mp } $, and backgrond fluids.\\
Equations (1) and (2) can be written as:
\begin{equation}
\rho _{u_{\mp }}=\pm \frac{1}{2}\dot{u}^{2}_{\mp } +V_{e}+V_{p}
\end{equation}
\begin{equation}
P _{u_{\mp }}=\pm \frac{1}{2}\dot{u}^{2}_{\mp } -V_{e}-V_{p}
\end{equation}
and according to the continuity equation of scalar field $ u_{\mp } $ :
\begin{equation}
\dot{\rho} _{u _{\mp }}+3H(\rho _{u _{\mp }}+p_{u_{\mp } })=0
\end{equation}
we have:
\[ \pm \dot{u}_{\mp }\ddot{u}_{\mp }+(V_{e}+V_{p})_{,\Psi } \dot{\Psi } \pm 3H\dot{u}^{2}_{\mp} =0   \]
where $ _{,\Psi } $ denotes a derivative with respect to the $ \Psi $. By substituting $ \dot{\Psi } $ as:
  \begin{equation}
  \dot{\Psi } =\frac{\dot{u} _{\mp }}{\sqrt{\mid 1-\gamma ^{2}\kappa ^{2}\Psi ^{2}(u_{\mp })\mid}} 
  \end{equation} 
the equation of motion of $ u_{\mp } $, can be found as:
\begin{equation}
  \ddot{u} _{\mp }+3H\dot{u}_{\mp} \pm \frac{V_{e,\Psi }+V_{p,\Psi}}{\sqrt{\mid 1-\gamma ^{2}\kappa ^{2}\Psi ^{2}(u_{\mp })\mid} }=0
  \end{equation}  
The continuity equation of background fluids also, can be written as:
\begin{equation}
\dot{\rho} _{bg}+3H(\rho _{bg }+p_{bg})=0
\end{equation}
so the evolution equations of different fluids, are:
\begin{equation}
\dot{\rho } _{r}+4H\rho _{r}=0
\end{equation}
\begin{equation}
\dot{\rho } _{d}+3H\rho _{d}=0
\end{equation}
\begin{equation}
\dot{\rho  } _{b}+3H\rho _{b}=0
\end{equation}
Also $ \frac{\dot{H} }{H^{2}} $ will be needed in order to find the evolution of dimensionless variables. From Eq. (6) we have:
\[ \frac{\dot{H} }{H^{2}}=\frac{\kappa ^{2}}{6H^{3}}\left[\pm \dot{u}_{\mp } \ddot{u}_{\mp} +\dot{\rho } _{r}+\dot{\rho } _{d}+\dot{\rho  } _{m}+(V_{e,\Psi }+V_{p,\Psi})\dot{\Psi }   \right]  \]
by substituting $ \ddot{u}_{\mp } $ from Eq. (15), $ \dot{\Psi } $ from Eq. (14), and $ \dot{\rho } _{r} $, $ \dot{\rho } _{d} $, $ \dot{\rho } _{m} $, from equations (17) to (19), and using equations (8) and (9), we have:
\begin{equation}
\frac{\dot{H} }{H^{2}}=-3+3(y^{2}_{e}+y^{2}_{p})+r^{2}+\frac{3}{2}(m^{2}_{d}+m^{2}_{b})
\end{equation}
Now, derivatives of dimensionless variables with respect to logarithm of scale factor, can be found as:
\begin{align*}
x'_{\mp }=&\frac{1}{H}\dot{x} _{\mp }=\frac{\kappa }{\sqrt{6}H}(\frac{\ddot{u}_{\mp }}{H}-\dot{u}_{\mp } \frac{\dot{H} }{H^{2}})\\
y'_{e}=& \frac{1}{H}\dot{y} _{e }=\frac{\kappa }{\sqrt{3}H}\sqrt{V_{,e}}(\frac{\dot{\Psi } }{2H}-\frac{\dot{H} }{H^{2}})\\
y'_{p}=& \frac{1}{H}\dot{y} _{p }=\frac{\kappa }{\sqrt{3}H}\sqrt{V_{,p}}(\frac{\dot{\Psi } }{2H}-\frac{\dot{H} }{H^{2}})\\
\rho '_{r}=& \frac{1}{H}\dot{\rho } _{r }=\frac{\kappa }{\sqrt{3}H}(\frac{\dot{\rho }_{r} }{H}-\rho_{r} \frac{\dot{H} }{H^{2}})\\
\rho '_{d}=& \frac{1}{H}\dot{\rho } _{d }=\frac{\kappa }{\sqrt{3}H}(\frac{\dot{\rho }_{d} }{H}-\rho_{d} \frac{\dot{H} }{H^{2}})\\
\rho '_{b}=& \frac{1}{H}\dot{\rho } _{b }=\frac{\kappa }{\sqrt{3}H}(\frac{\dot{\rho }_{b} }{H}-\rho_{b} \frac{\dot{H} }{H^{2}})\\
\end{align*}
by substituting $ \ddot{u} $ from Eq. (15), $ \dot{\Psi } $ from Eq. (14), $ \frac{\dot{H} }{H^{2}} $ from Eq. (20),  and $ \dot{\rho } _{r} $, $ \dot{\rho } _{d} $, $ \dot{\rho } _{m} $, from equations (17) to (19), and using equations (8) and (9), the relations above can be written as:\\
\begin{align}
 x^{'}_{\mp }&=-x_{\mp }[3(y^{2}_{e}+y^{2}_{p})+r^{2}+\frac{3}{2}(m^{2}_{d}+m^{2}_{d})]+ \sqrt{\frac{3}{2}}\ \frac{\mp \alpha y_{p}^{2}\pm \beta (\kappa \Psi )^{\beta }y_{e}^{2}}{\kappa  \Psi \sqrt{\lvert1-\gamma ^{2}\kappa ^{2}\Psi ^{2}(u_{\mp }) \lvert }}\\
y_{p}^{'}&=y_{p}[3-3(y^{2}_{e}+y^{2}_{p})-r^{2}-\frac{3}{2}(m^{2}_{d}+m^{2}_{d})]+\sqrt{\frac{3}{2}}\ \frac{\alpha y_{p}x_{\mp }}{ \kappa \Psi \sqrt{\lvert1-\gamma ^{2}\kappa ^{2}\Psi ^{2}(u_{\mp }) \lvert }}
\end{align}\\
\begin{align}
y_{e}^{'}&=y_{e}[3-3(y^{2}_{e}+y^{2}_{p})-r^{2}-\frac{3}{2}(m^{2}_{d}+m^{2}_{d})]-\sqrt{\frac{3}{2}}\ \frac{\beta (\kappa \Psi )^{\beta }y_{e}x_{\mp }}{ \kappa \Psi \sqrt{\lvert1-\gamma ^{2}\kappa ^{2}\Psi ^{2}(u_{\mp }) \lvert }}\\
r^{'}&=r[1-3(y^{2}_{e}+y^{2}_{p})-r^{2}-\frac{3}{2}(m^{2}_{d}+m^{2}_{d})]\\
m_{d}^{'}&=m_{d}[\frac{3}{2}-3(y^{2}_{e}+y^{2}_{p})-r^{2}-\frac{3}{2}(m^{2}_{d}+m^{2}_{d})]\\
m_{b}^{'}&=m_{b}[\frac{3}{2}-3(y^{2}_{e}+y^{2}_{p})-r^{2}-\frac{3}{2}(m^{2}_{d}+m^{2}_{d})]
\end{align}\\
Equations (21) to (26), make a non-autonomous system of equations; but by adding the equation:
\begin{equation}
\Psi^{'} =\frac{\sqrt{6}x_{\mp }}{\kappa \sqrt{\lvert1-\gamma ^{2}\kappa ^{2}\Psi ^{2}(u_{\mp }) \lvert}}
\end{equation}
to the system, it will change to the autonomous system of seven equations which define the evolution of dimensionless variables.\\
Numerical solution of this system can be found starting from a very early point dominated by radiation ($ r_{initial}= 1 $ ) and the other fluids are negligible. Figure (1) shows the evolution of density parameters found from numerical solution of autonomous system, where $ \alpha  = 3 $, $ \beta  = 2 $ and $ \gamma =0.01 $; this picture is compatible with our universe; and according to this picture, 
starting from radiation domination, exponential (or gaussian) potential appears later in our early universe and causes inflation; then the kinetic energy of scalar matter field dominates our universe and with its positive pressure slows down the expansion of the universe, remarkably; then radiation dominates the universe for second time and after that era, dark matter and baryonic matter appears in our universe; but power-law potential appears later in our universe and its today's density parameter is the same as dark energy density of today's universe [5]. power-law potential will dominate our universe in the future. According to the numerical solution, today’s values of energy density parameters are:\\
\begin{align*}
\Omega _{\Lambda,0 }&=y_{p,0}^{2}=0.683\\
\Omega _{m,0}&=m_{d,0}^{2}= 0.267\\
\Omega _{b,0}&=m_{b,0}^{2}= 0.0489\\
\Omega _{r ,0}&=r_{0}^{2} \ \ = 8.28\times 10^{-5}
\end{align*}
these values are very close to the measured ones [5].\\
\begin{figure}
  \begin{center}
    \includegraphics[width=1.0\textwidth]{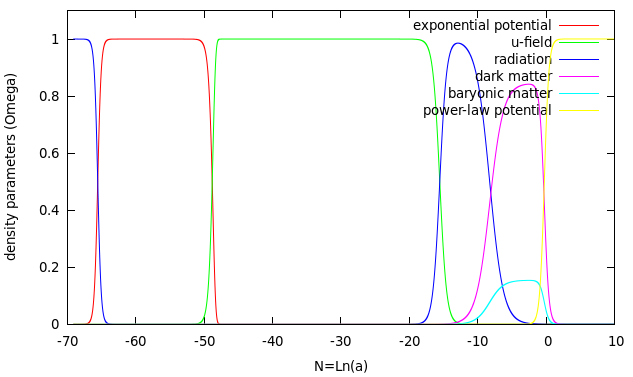}
    \label{fig:}
    \caption{\small Evolution of energy density parameters for  $ \alpha =3 $,  $ \beta  =2 $, $ \gamma =0.01 $ ; starting from a point dominated by radiation, $ r_{init.}=1 $ . Dark energy, dark Matter and baryonic matter density parameters of today's universe are:  $ \Omega _{\Lambda }\simeq 0.683 $ , $ \Omega _{d}\simeq 0.267 $ and $ \Omega _{b}\simeq 0.0489 $.}
  \end{center}
\end{figure}
\subsection{\small HUBBLE PARAMETER}
The values of Hubble parameter, over time, can be found numerically, by adding the equation below to the system of autonomous equations:
\begin{equation}
H'=\frac{\dot{H} }{H}=H[-3+3(y^{2}_{e}+y^{2}_{p})+r^{2}+\frac{3}{2}(m^{2}_{d}+m^{2}_{b})]
\end{equation}
According to the picture (2), Hubble parameter of very early universe was very high. When positive pressure of scalar matter field and then radiation, dominates our universe, Hubble parameter decreases; but because of the negative pressure of power-law potential, it’s going to increase in the future. Figure (3) shows that how equation of state (so the pressure) of our universe changes over time.\\
Today’s value of Hubble parameter found from numerical solution is:
\[ H_{0}=2.18 \times 10^{-18} \ s^{-1}\simeq 67.36 \ km/s/Mpc  \]
that is the same as its observed value [5].
      \begin{figure}
       \includegraphics[width=1.0\textwidth]{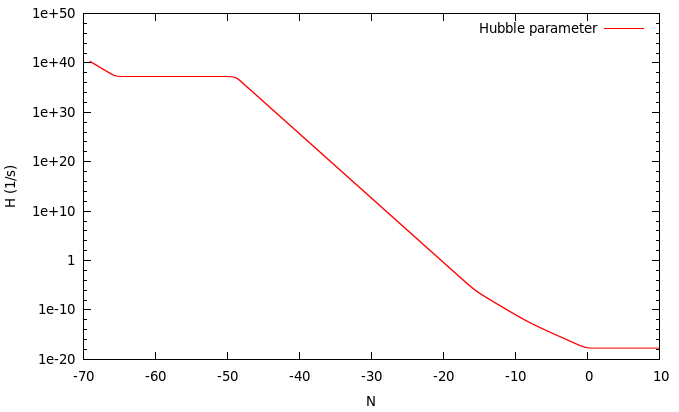}
      \label{fig:}
      \caption{\small Evolution of Hubble parameter for  $ \alpha =3 $,  $ \beta  =2 $, $ \gamma =0.01 $, when our universe starts from radiation dominated era, $ r_{initial}=1 $. Today's value of Hubble parameter for this case is: $H_{0}=2.18 \times 10^{-18} \ s^{-1}\simeq 67.36 \ km/s/Mpc  $. }
      \end{figure}
\begin{figure}
      \includegraphics[width=1.0\textwidth]{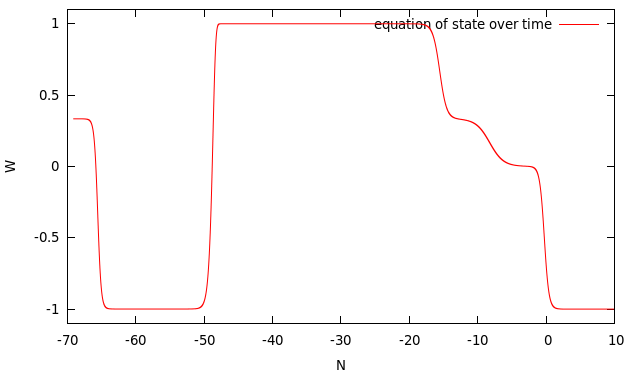}
      \label{fig:}
      \caption{\small Evolution of equation of state of our universe, for  $ \alpha =3 $,  $ \beta  =2 $, $ \gamma =0.01 $, when our universe starts from radiation dominated era, $ r_{initial}=1 $ . }
        \end{figure} \\

\subsection{\small SIZE AND AGE OF THE UNIVERSE}
\ The size of the universe at any time, called conformal time $ \eta(t)  $, is given by:
\[ \eta (t)=\int_{0}^{t} \frac{cdt^{'}}{a(t^{'})}\ \ \ \ or\ \ \ \frac{d\eta }{dt}=\frac{c}{a} \]
\begin{equation}
\Longrightarrow \ \ \ \eta ^{'}=\frac{d\eta }{dN}=\frac{1}{H}\frac{c}{a}
\end{equation}
\
On the other hand, the proper distance of particle horizon of a universe starting from a Big Bang, is given by:
\[ d_{p}^{PH}(t)=a(t)\int_{0}^{t} \frac{cdt^{'}}{a(t^{'})}\ \ \ \ \Longrightarrow \ \ \ \frac{d}{dt}[   \frac{d_{p}^{PH}(t)}{a(t)}]=\frac{c}{a(t)} \]
\begin{equation}
\Longrightarrow \ \ \ \frac{d}{dN}[d_{p}^{PH}(t) ]=d_{p}^{PH}(t) +\frac{c}{H}
\end{equation}
\
$d_{p}^{PH}(t)  $ is in fact the size of our today's visible universe, at any time in the past; but $ \eta(t)  $ ( or comoving proper distance) is the size of particle horizon at any time. \\
By adding equations (29) and (30) to the autonomous system of equations, the numerical solutions of these equations can be found over time, as shown in figure (4). According to the figure, when the horizon of the inflationary universe is about 1 meter, the size of our visible universe is negligible; and when the size of our visible universe is about 1 meter, the particle horizon is about $ 10^{15} $ meters; the age of the universe is about $ 10^{-7} $ seconds at that time. The size and age of the today's universe according to the picture, are:\\
\[ \ \ \ \ \ \  \eta _{0}=4.36\times 10^{26}\ m \simeq 14.141 \ \ Giga\ parsecs \]
\[ t_{0}=4.349\times 10^{17} \ s\simeq 13.7 \ Giga\ year \]
they are very close to the observed values [5].
\begin{figure} 
       \includegraphics[width=1.0\textwidth]{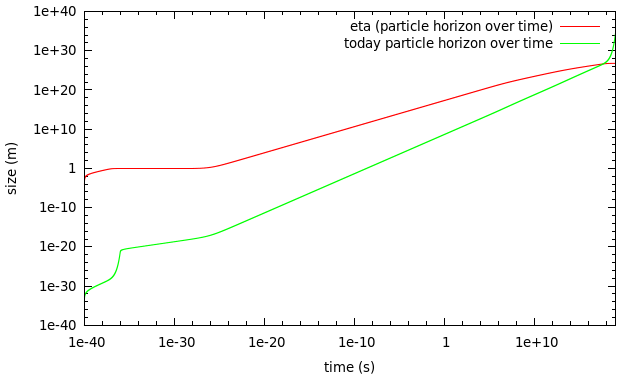}
      \label{fig:}
      \caption{\small Evolution of conformal time and today’s particle horizon over time for  $ \alpha =3 $,  $ \beta  =2 $, $ \gamma =0.01 $, when our universe starts from radiation dominated era, $ r_{initial}=1 $ . The size of today’s universe is about $ 4.36\times 10^{26}\ m \simeq 14.141 \ \ Giga\ parsecs$ ; and its age is about $ 4.349\times 10^{17} \ s\simeq 13.7 \ Giga\ year $. }
      \end{figure} \\

\section{\small CONCLUSIONS}
In order to find an explanation for early inflationary universe and accelerated expansion of today's universe, I've defined a hybrid potential of scalar matter field $ \Psi  $, that is a combination of a power-law potential and an exponential potential (gaussian) of $ \Psi  $-field. The exponential potential energy dominates very early universe and it can explain inflation, and power-law potential energy is a good alternative for dark energy. \\
The results found from the numerical solution of the system of equations of dimensionless variables defining the evolution of a universe filled with background fluids (dark matter, baryonic matter and radiation) and scalar matter field $ \Psi  $ (with hybrid potential), are consistent with real universe. The plot of energy density parameters gives the real history of our universe; and today's values of density parameters of dark energy, dark matter, baryonic matter, and Hubble parameter, and the age and size of our universe, found from this model, are very close to (and some times the same as) measured values.\\ 
\
\newline
\textbf{\Large  References}\\
\newline
$ [1] $ Paul J. Steinhardt. A quintessential introduction to dark energy.
Department of Physics, Princeton University, Princeton, NJ 08540, USA.\\
Published online 17 September 2003.\\
\newline
$ [2] $ A.D. Dolgov. PROBLEMS OF VACUUM ENERGY AND DARK ENERGY.
 arXiv:hep-ph/0405089v1 11 May 2004.\\
\newline
$ [3] $ Caldwell, R. R., Dave, R. and Steinhardt, P.J. Cosmological imprint of an energy component with general equation of state.
Phys. Rev. Lett. 80, 1582-1585, 1998.\\
\newline
$ [4] $ Kari Enqvist, Tomi Koivisto, Gerasimos Rigopoulos. Non-metric chaotic inflation.
ArXiv:1107.3739v1, 9 Jul.2011.\\
\newline
$ [5] $ Planck Collaboration. Planck 2013 results. I. Overview of products and scientific results.
 Arxiv:1303.5062v1, 20 Mar.2013.\\

 \end{document}